\documentclass[aps,prd,nofootinbib,superscriptaddress,twocolumn]{revtex4}  
\usepackage{graphicx}
\usepackage{epstopdf}
\usepackage{amsmath}
\usepackage{amsfonts}
\usepackage{amssymb}
\usepackage{appendix}
\usepackage{comment}
\usepackage{bbold}
\usepackage{color}
\usepackage{slashed}
\usepackage{subfigure}
\usepackage{setspace}
\usepackage{footnote}
\usepackage{multirow}
\usepackage{mathrsfs}
\usepackage{feynmp-auto}
\usepackage{hyperref}
\hypersetup{colorlinks,citecolor= nicegreen,linkcolor= nicered}
\definecolor{niceblue}{rgb}{0.388235, 0.627451, 0.847059}
\definecolor{nicered}{rgb}{0.7,0.1,0.1}
\definecolor{nicegreen}{rgb}{0.1,0.5,0.1}

\newcommand{\be}{\begin{equation}}
\newcommand{\ee}{\end{equation}}
\newcommand{\ba}{\begin{array}}
\newcommand{\ea}{\end{array}}
\newcommand{\bea}{\begin{eqnarray}}
\newcommand{\eea}{\end{eqnarray}}
\newcommand{\balg}{\begin{align}}
\newcommand{\ealg}{\end{align}}
\newcommand{\bit}{\begin{itemize}}
\newcommand{\eit}{\end{itemize}}
\newcommand{\trm}[1]{\textrm{#1}}

\newcommand{\Mpc}{\trm{\Mpc}}
\newcommand{\yr}{\trm{\yr}}
\newcommand{\eV}{\trm{\eV}}

\newcommand{\NMplane}{$\Delta N_{\rm eff}$ -- $m^{\rm eff}_{\rm sterile} \,$}

\begin{document}

\singlespacing
\allowdisplaybreaks

\title{Constraining Sterile Neutrino Cosmology with Terrestrial Oscillation Experiments}

\author{Jeffrey M. Berryman}
\affiliation{Center for Neutrino Physics, Department of Physics, Virginia Tech, Blacksburg, VA 24061, USA}


\begin{abstract}
We explore the complementarity between terrestrial neutrino oscillation experiments and astrophysical/cosmological measurements in probing the existence of sterile neutrinos. We find that upcoming accelerator neutrino experiments will not improve on constraints by the time they are operational, but that reactor experiments can already probe parameter space beyond the reach of Planck. We emphasize the tension between cosmological experiments and reactor antineutrino experiments and enumerate several possibilities for resolving this tension.
\end{abstract}

\maketitle

\section{Introduction}
\label{sec:intro}
\setcounter{equation}{0}

As the precision of neutrino experiments continues to improve, the importance of interdisciplinary studies grows --- it becomes increasingly possible and imperative to determine the extent to which the three-neutrino oscillation framework can simultaneously describe a broad set of experimental results. In many ways, neutrino oscillations are already an interdisciplinary endeavor: neutrinos of different origins (solar, reactor, accelerator, atmospheric, cosmogenic, etc.) over orders of magnitude of energy have been studied in a common framework. As a result, synergies and tensions \cite{Esteban:2018azc} have arisen between different sectors and between experiments within the same sector. Understanding the origins of these tensions is a central issue.

One proposed solution to some (but not all) of these tensions is that additional species of neutrinos with eV-scale masses exist. The evidence driving this derives from anomalous measurements of electron-neutrino disappearance \cite{Acero:2007su,Giunti:2010zu,Mention:2011rk,Hayes:2013wra} and electron-neutrino appearance \cite{Aguilar:2001ty,Aguilar-Arevalo:2013pmq} (see, for instance, Refs.~\cite{Gariazzo:2017fdh,Giunti:2017yid,Dentler:2017tkw,Gariazzo:2018mwd,Dentler:2018sju,Giunti:2019qlt,Giunti:2019aiy} for more details). It is important to note that this is not a silver-bullet solution to these anomalies; even among the subset of anomalous experimental results, a consistent description in terms of additional neutrinos is lacking, as has been explored in, for example, Ref.~\cite{Dentler:2018sju}.

In addition to the overall consistency of dedicated oscillation experiments, astrophysics and cosmology present stiff challenges to the proposed existence of additional neutrinos. The literature concerning the role of sterile neutrinos\footnote{There is no reason to suspect, \emph{a priori}, that additional neutrinos do not possess separate interactions among themselves or with other particle species. We will assume that additional neutrinos have no interactions, but use the term ``sterile'' to refer to any species that does not have weak interactions.} in cosmology is vast; see Refs.~\cite{Melchiorri:2008gq,Archidiacono:2012ri,Archidiacono:2013xxa,Mirizzi:2013gnd,Gariazzo:2013gua,Archidiacono:2014apa,Bridle:2016isd,Ghalsasi:2016pcj,Knee:2018rvj,Kang:2019xuq} and references therein. Introducing these additional, light degrees of freedom to resolve oscillation anomalies necessarily invokes constraints from an ostensibly disconnected field of study. The objectives are to (a) understand how different sets of constraints complement each other, and (b) determine the essential characteristics of a solution to all sets of constraints.

In this work, we focus on the first of these two objectives. Our goals are to determine how constraints on sterile neutrinos derived from terrestrial oscillation experiments translate into constraints on their cosmological properties and whether or not these improve on constraints from dedicated astrophysics/cosmology experiments. The focus will be on accelerator and reactor (anti)neutrino experiments. Additionally, the observation of coherent elastic neutrino-nucleus scattering (CE$\nu$NS) by the COHERENT collaboration \cite{Akimov:2017ade} has opened a new means by which to study neutrinos and their interactions \cite{Anderson:2012pn,Dutta:2015nlo,Kerman:2016jqp,Canas:2017umu,Kosmas:2017tsq}. We will consider the sensitivities of several proposed experiments that rely on this process which exist to test the sterile-neutrino interpretation of reactor antineutrino anomalies.

This manuscript is organized as follows. In Sec.~\ref{sec:formalism}, we review oscillations involving a sterile neutrino and the terrestrial constraints considered are discussed in Sec.~\ref{sec:experiments}. The framework with which we study the cosmology of a sterile neutrino is presented in Sec.~\ref{sec:cosmology} and our results are given in Sec.~\ref{sec:results}. We offer concluding thoughts in Sec.~\ref{sec:conclusions}.


\section{Oscillations with Sterile Neutrinos}
\label{sec:formalism}
\setcounter{equation}{0}

Oscillations with three neutrinos can be extended to include a fourth neutrino by generalizing the $3\times3$ Pontecorvo-Maki-Nakagawa-Sakata (PMNS) matrix to a $4 \times 4$, unitary matrix, $U$, whose matrix elements we denote $U_{\alpha i}$ with $\alpha = e, \, \mu, \, \tau, \, s$ and $i = 1, \, 2, \, 3, \, 4$. We need not concern ourselves with the parametrization of the entire matrix (see, for instance, Ref.~\cite{Berryman:2015nua} for more details); the relevant elements of this matrix, for this study, are
\begin{eqnarray}
|U_{e4}|^2 & = & s^2_{14},  \\
|U_{\mu4}|^2 & = & s^2_{24} c^2_{14},
\end{eqnarray}
where we define $s_{ij} \equiv \sin \phi_{ij}$ and $c_{ij} \equiv \cos \phi_{ij}$, where $\phi_{ij}$ are the mixing angles that parametrize $U$.

The probability that a neutrino of initial flavor $\alpha$ ($= e, \, \mu, \, \tau, \, s$) propagating in vacuum will be detected\footnote{It is, by hypothesis, not possible to directly detect sterile neutrino states, as they do not interact with detectors. Instead, one must infer that neutrinos have converted into the sterile flavor by observing a reduced interaction rate.} with flavor $\beta$ ($= e, \, \mu, \, \tau, \, s$) is
\begin{align}
\label{eq:osc_probs}
P_{\alpha\beta} = & \, \,  \big| \delta_{\alpha\beta} - U_{\alpha 2} U^*_{\beta 2} \left(1 - e^{-i \Delta_{21}} \right) - U_{\alpha 3} U^*_{\beta 3} \left(1 - e^{-i \Delta_{31}} \right) \big. \nonumber \\
& - \big. U_{\alpha 4} U^*_{\beta 4} \left(1 - e^{-i \Delta_{41}} \right) \big|^2,
\end{align}
with $\Delta_{ij} \equiv 2.54 \left( \Delta m_{ij}^2 / \text{eV}^2 \right) \left( L /\text{km} \right) \left( \text{GeV} / E_\nu \right)$, where $\Delta m^2_{ij} \equiv m^2_i - m^2_j$ is the difference in neutrino masses squared, $L$ is the distance traveled and $E_\nu$ is the energy. In the limit $\Delta_{31}, \Delta_{21} \ll 1$, Eq.~\eqref{eq:osc_probs} simplifies to
\begin{align}
\label{eq:osc_probs2}
P_{\alpha \beta} & \approx \big| \delta_{\alpha \beta} - U_{\alpha 4} U^*_{\beta 4} \left(1 - e^{-i \Delta_{41}} \right) \big|^2 \nonumber \\
& = \left\{ \begin{array}{c} 
1 - \sin^2 2\theta_{\alpha\alpha} \sin^2 \frac{\Delta_{41}}{2} \quad (\alpha = \beta) \nonumber \\ 
\sin^2 2\theta_{\alpha \beta} \sin^2 \frac{\Delta_{41}}{2} \quad (\alpha \neq \beta) 
\end{array}  \right. ,
\end{align}
where have employed the effective mixing angles
\begin{eqnarray}
\sin^2 2\theta_{\alpha \alpha} & \equiv & 4 |U_{\alpha 4}|^2 (1-|U_{\alpha 4}|^2), \label{eq:thaa}\\
\sin^2 2\theta_{\alpha \beta} & \equiv & 4 |U_{\alpha 4}|^2 |U_{\beta 4}|^2 \label{eq:thab}.
\end{eqnarray}
The quantities $\sin^2 2\theta_{ee}$ and $\sin^2 2\theta_{\mu\mu}$, as well as the mass-squared splitting $\Delta m^2_{41}$, are the primary subjects of this work.

We are interested in scenarios in which only one of the active-sterile mixing angles is nonzero; in particular, we are concerned with either nonzero $\phi_{14} \, ( = \theta_{ee})$ or $\phi_{24} \, ( = \theta_{\mu\mu})$.\footnote{The first relation is always true, while the second is only true if $\phi_{14}$ vanishes.} The reason for this is that the formalism we introduce in Sec.~\ref{sec:cosmology} relies on the two-flavor-oscillations approximation in the early Universe. This approximation is violated if two active-sterile mixing angles are simultaneously nonzero; our framework allows for us to translate bounds on $\sin^2 2\theta_{ee}$ and $\sin^2 2\theta_{\mu\mu}$ into cosmology parameter space, but not bounds on $\sin^2 2\theta_{e\mu}$. In other words, we study neutrino \emph{disappearance} anomalies, but not neutrino \emph{appearance} anomalies.

When neutrinos propagate through matter, their propagation is altered by a background matter potential \cite{Wolfenstein:1977ue}. The propagation Hamiltonian receives an additional contribution that is diagonal in the flavor basis, $(\delta H)_{\alpha \beta} = \sqrt{2} G_F \times \text{diag}\left( n_e, 0, 0, \frac{1}{2} n_n \right) $, where $n_e$ and $n_n$ are, respectively, the electron and neutron number densities along the path of propagation. In the mass basis, this potential becomes $(\delta H)_{ij} = U^*_{\alpha i} (\delta H)_{\alpha \beta} U_{\beta j}$; it is no longer diagonal, rendering the mass-basis propagation Hamiltonian similarly nondiagonal. For antineutrinos, we must replace $(\delta H)_{\alpha \beta} \to - (\delta H)_{\alpha \beta}$, and $U \leftrightarrow U^*$. 

Two key assumptions in writing down this matter potential are (i) that interactions of neutrinos with the background matter are forward, coherent and elastic, and (ii) that neutrinos do not interact with the potential produced by other neutrinos. There are physical systems in which these assumptions do not hold -- for instance, neutrinos propagating out of supernovae \cite{Mirizzi:2015eza} -- and the resulting phenomenon is referred to as collective oscillations \cite{Duan:2010bg}. In Sec.~\ref{sec:cosmology}, we discuss another such system: neutrinos in the early Universe.


\section{Terrestrial Constraints on a Fourth Neutrino}
\label{sec:experiments}
\setcounter{equation}{0}

We discuss the neutrino experiments that we use to constrain sterile neutrino cosmology. There is a great deal of activity in the literature regarding searches for a fourth neutrino (see, for instance, Ref.~\cite{Abazajian:2012ys,Kopp:2013vaa,Gariazzo:2017fdh,Dentler:2018sju}, and references therein), but we have chosen a subset that we believe provides the most value in terms of illuminating the connection between terrestrial neutrino oscillation experiments and early-Universe cosmology.

\subsection*{Accelerator Neutrino Experiments}

Neutrinos for the MINOS experiment \cite{Adamson:2013whj,Adamson:2014vgd} are produced at the Fermilab Main Injector by protons incident on a graphite target; they are directed towards a near detector and a far detector that are 1.4 km and 735 km away, respectively. We show in red the 90\% C.L. constraint derived in Ref.~\cite{Adamson:2017uda}.

\begin{figure}[!tbp]
\centerline{\includegraphics[width=\linewidth]{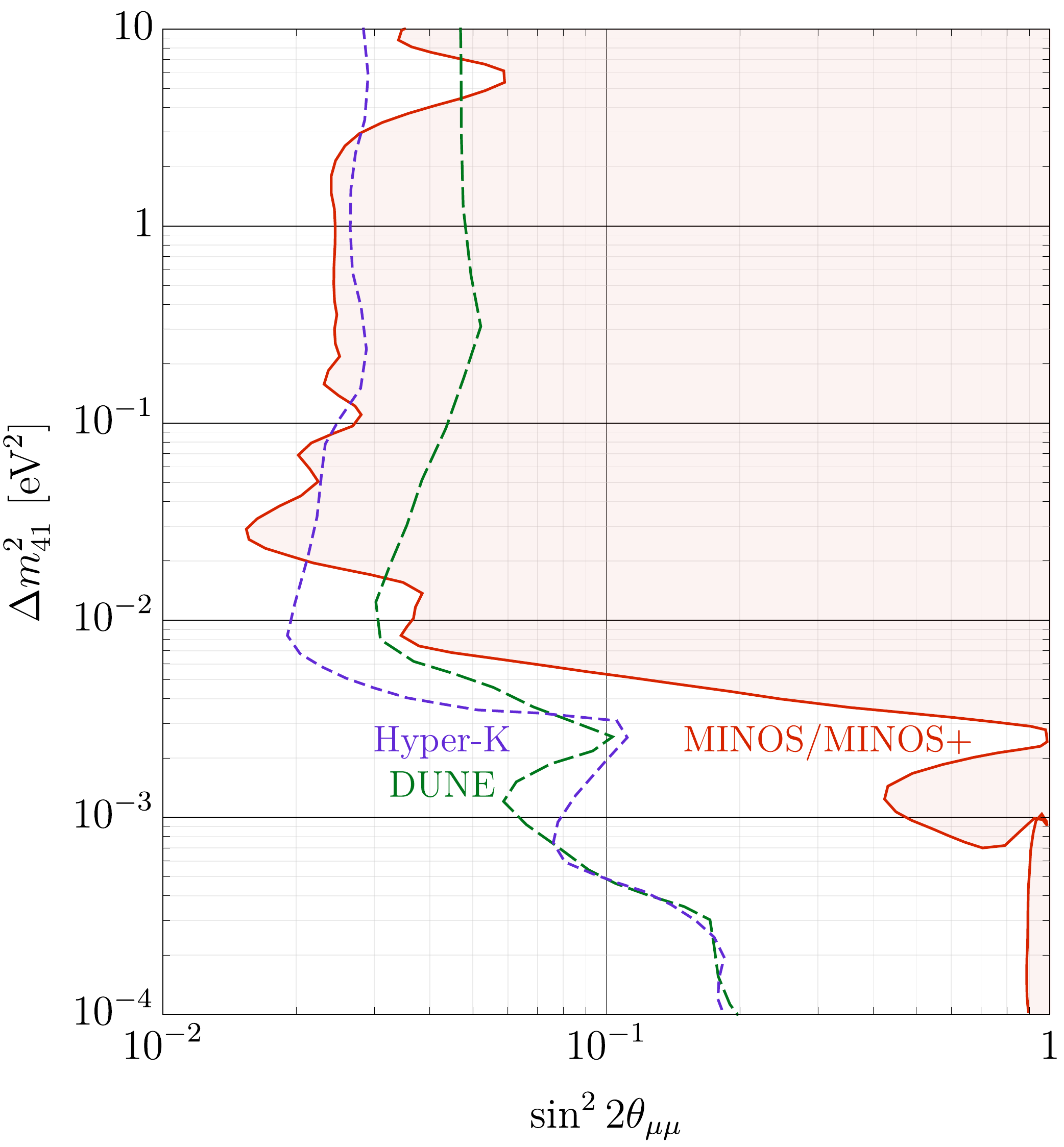}}
\caption{Constraints in the $\sin^2 2\theta_{\mu\mu}$--$\Delta m_{41}^2$ plane from accelerator neutrino experiments. The red, shaded region is excluded at 90\% C.L. by MINOS/MINOS+. Also shown are the expected 95\% C.L. sensitivities of DUNE (long-dashed green) and Hyper-K (short-dashed purple).}
\label{fig:accelerator_data}
\end{figure}

We also show the sensitivities of two proposed, long-baseline accelerator neutrino experiments, estimated using Monte Carlo simulations. The Deep Underground Neutrino Experiment (DUNE) \cite{Adams:2013qkq,Acciarri:2015uup} is a proposed 34-kton liquid argon detector located 1300 km from the Fermilab main injector.  The Hyper-Kamiokande (Hyper-K) experiment \cite{Abe:2015zbg,Abe:2016ero,Abe:2018uyc} consists of two water Cerenkov detectors with a combined fiducial mass of 0.56 Mton located 295 km from the Japan Proton Accelerator Research Complex (J-PARC). We discuss our simulations of these experiments in Appendix \ref{app:DUNE}. The resulting 95\% C.L. sensitivities for DUNE (long-dashed green) and Hyper-K (short-dashed purple) in the $\sin^2 2 \theta_{\mu\mu}$ -- $\Delta m^2_{41}$ plane are shown in Figure~\ref{fig:accelerator_data}.
 
\subsection*{Reactor Antineutrino Experiments}

\begin{figure}[!t]
\centerline{\includegraphics[width=\linewidth]{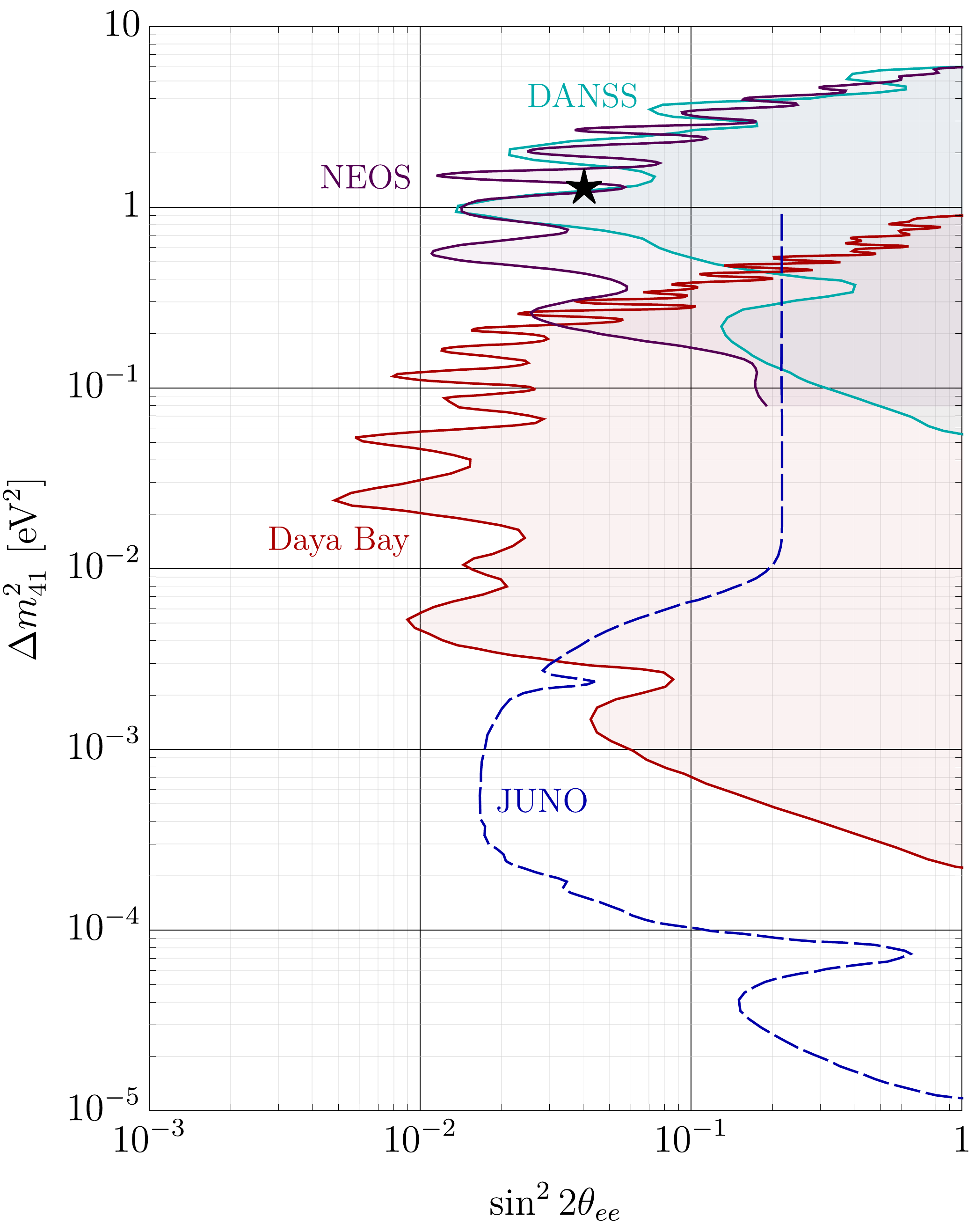}}
\caption{Constraints in the $\sin^2 2\theta_{ee}$--$\Delta m_{41}^2$ plane from the reactor antineutrino experiments. The shaded regions are currently excluded by DANSS (cyan, 95\% C.L.), Daya Bay (red, 95\% C.L.), NEOS (purple, 90\% C.L.). Also shown is the sensitivity reach of JUNO (long-dashed blue, 95\% C.L.). The black, five-pointed start represents the best-fit point from the global analysis of Ref.~\cite{Dentler:2018sju}.}
\label{fig:reactor_data}
\end{figure}

The Detector of AntiNeutrino based on Solid Scintillator (DANSS) project \cite{Danilov:2014vra,Alekseev:2016llm,Alekseev:2018efk} consists of an array of gadolinium-coated plastic scintillators in an experimental hall beneath a reactor at the Kalinin Nuclear Power Plant. The detector is placed in a lifting system that allows the distance between the reactor and the detector to vary between 9.7 m and 12.2 m. The cyan curve in Figure \ref{fig:reactor_data} shows the sensitivity (95\% C.L.) of DANSS to the presence of a sterile neutrino as presented in Ref.~\cite{Alekseev:2018efk}.

The Daya Bay experiment \cite{An:2016ses} is a collection of eight antineutrino gadolinium-doped liquid scintillator antineutrino detectors observing the antineutrinos produced by six reactor cores located in southern China. The sensitivity of Daya Bay to a sterile neutrino was investigated in Ref.~\cite{An:2016luf}, and the resulting exclusion limit (95\% C.L.) in the $\sin^2 2\theta_{ee}$ -- $\Delta m^2_{41}$ plane is reproduced in red in Figure \ref{fig:reactor_data}. An updated data set has recently been released \cite{Adey:2018zwh,Adey:2019ywk}, but a sterile-neutrino analysis has not yet been performed.

The NEOS experiment \cite{Ko:2016owz} consists of a single gadolinium-doped liquid scintillator detector located 24 m from a reactor core in southwestern South Korea.  A sterile neutrino search was performed in Ref.~\cite{Ko:2016owz}, and the resulting limit (90\% C.L.) is shown in purple in Fig.~\ref{fig:reactor_data}.

The Jiangmen Underground Neutrino Observatory (JUNO) experiment \cite{An:2015jdp} is a proposed 20-kton liquid scintillator detector whose primary physics goal is to determine the neutrino mass hierarchy. The sensitivity of JUNO to the presence of sterile neutrinos was also studied in Ref.~\cite{An:2015jdp}; the result (95\% C.L.) is shown in long-dashed blue in Fig.~\ref{fig:reactor_data} assuming a normal hierarchy of both active and sterile neutrino masses, i.e., assuming the neutrino masses are ordered $m_1<m_3$ and $m_1<m_4$.

The next few years will see a the number of new short-baseline results from STEREO \cite{Almazan:2018wln}, SoLid \cite{Abreu:2018pxg} and PROSPECT \cite{Ashenfelter:2018iov}, the first and last of which have already started collecting data. However, to avoid clutter in Fig.~\ref{fig:reactor_data} and in our results in Sec.~\ref{sec:results}, we do not consider these in our analysis. While they will ultimately be competitive in the $\Delta m_{41}^2 \sim \mathcal{O}(1-10)$ eV$^2$ region, they have not yet produced world-leading limits. Constraints on sterile neutrinos mixing with $\nu_e$/$\overline{\nu}_e$ from radioactive source \cite{Hampel:1997fc,Abdurashitov:1998ne,Abdurashitov:2005tb}, solar \cite{Cleveland:1998nv,Aharmim:2005gt,Hosaka:2005um,Aharmim:2006kv,Cravens:2008aa,Aharmim:2008kc,Bellini:2008mr,Abdurashitov:2009tn,Kaether:2010ag,Abe:2010hy,Bellini:2011rx,Gando:2014wjd} and carbon-scattering \cite{Auerbach:2001hz,Conrad:2011ce} experiments can be nontrivial, but we do not consider these.

A global fit of the 3+1 neutrino oscillation framework to short-baseline oscillation data was recently performed in Ref.~\cite{Dentler:2018sju}. The authors find that the best-fit point to global $\overline{\nu}_e$ disappearance data -- including reactor, gallium, solar and $\beta$-decay experiments -- with unfixed reactor fluxes $\{ \Delta m_{41}^2, \, \sin^2 2\theta_{ee}\} = \{ 1.3 \text{ eV}^2, \, 4.04 \times 10^{-2} \}$. This point is indicated by the black star in Fig.~\ref{fig:reactor_data}.
 
\subsection*{Low-Threshold Experiments}

The possibility of observing active-sterile oscillations through coherent elastic neutrino-nucleus scattering (CE$\nu$NS) has been previously discussed in Refs.~\cite{Anderson:2012pn,Dutta:2015nlo,Kerman:2016jqp,Canas:2017umu,Kosmas:2017tsq}; these provide a complementary probe of the $\sin^2 2\theta_{ee}$--$\Delta m_{41}^2$ space to reactor antineutrino experiments. In fact, the experiments that we will consider are also based at nuclear reactors, but we separate these from those of the previous subsection because the underlying signal process -- CE$\nu$NS -- is distinct from inverse beta decay (IBD).

\begin{figure}[!t]
\centerline{\includegraphics[width=\linewidth]{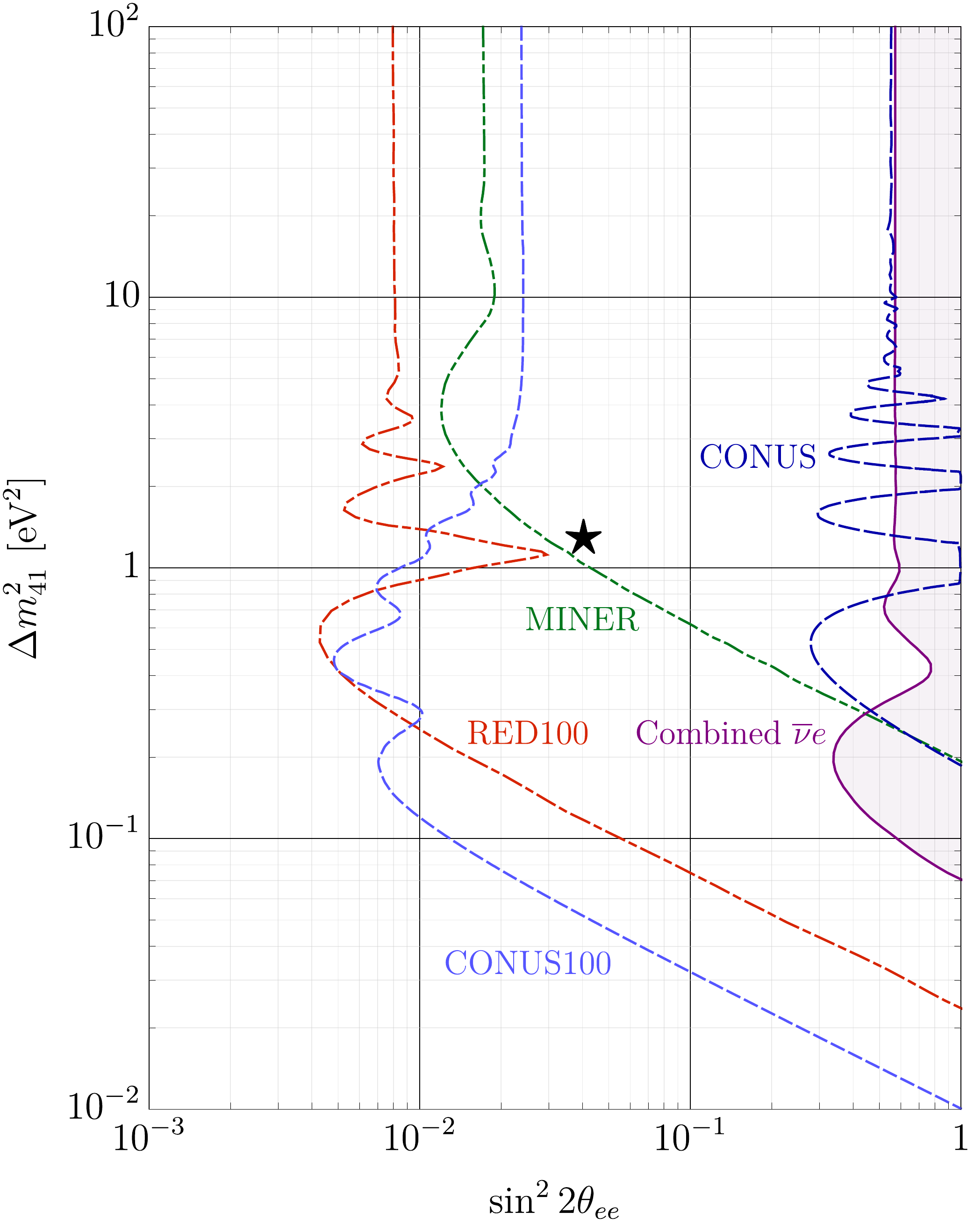}}
\caption{Constraints in the $\sin^2 2\theta_{ee}$--$\Delta m_{41}^2$ plane from low-threshold experiments. The purple, shaded region is excluded at 90\% C.L. by $\overline{\nu}e$ scattering \cite{Canas:2017umu}. Also shown are the expected sensitivities of RED100 (dot-dashed red, 90\% C.L.), MINER (double-dot-dashed green, 90\% C.L.), CONUS (long-dashed blue, 95\% C.L.) and CONUS100 (short-dashed light blue, 95\% C.L.). The black, five-pointed start represents the best-fit point from the global analysis of Ref.~\cite{Dentler:2018sju}.}
\label{fig:CENNS}
\end{figure}

Ref.~\cite{Canas:2017umu} reports a constraint in the $\sin^2 2\theta_{ee}$--$\Delta m_{41}^2$ plane from a combined analysis of $\overline{\nu}e$ scattering data from the Krasnoyarsk \cite{Vidyakin:1992nf}, Rovno \cite{Derbin:1993wy}, MUNU \cite{Amsler:1997pn} and TEXONO \cite{Deniz:2009mu} experiments.\footnote{This is, of course, not a CE$\nu$NS process, which is why we've opted to call this class of bounds ``low threshold."} We reproduce the resulting exclusion in purple in Fig.~\ref{fig:CENNS}. This constraint is quite weak relative to the other experiments in the figure, but it is the only such analysis that can currently exclude any portion of this parameter space.

We focus on a subset the many (existing and proposed) CE$\nu$NS experiments \cite{Wong:2008vk,Beda:2013mta,Gutlein:2014gma,Belov:2015ufh,Billard:2016giu,Aguilar-Arevalo:2016qen,Strauss:2017cuu,Kosmas:2017zbh}, starting with the RED100 \cite{Akimov:2012aya} and MINER \cite{Agnolet:2016zir} proposals. The sensitivities of these experiments to sterile neutrinos were studied in Ref.~\cite{Canas:2017umu}. Several benchmark scenarios were considered for each of these experiments; we consider the most aggressive scenarios to assess how these experiments fare under the most optimistic assumptions. We take baselines of 15 m for RED100 and 1 m for MINER and assume a 100\% efficiency for each. The 90\% C.L. sensitivities for RED100 and MINER are reproduced in dot-dashed red and double-dot-dashed green, respectively, in Figure~\ref{fig:CENNS}.

The sensitivity of the COHERENT experiment \cite{Akimov:2017ade} to sterile neutrinos has been studied in Refs.~\cite{Kosmas:2017tsq,Kosmas:2017zbh}. However, COHERENT is more sensitive to a sterile neutrino mixing with $\nu_\mu$/$\overline{\nu}_\mu$ than it is to mixing with $\nu_e$/$\overline{\nu}_e$. Consequently, we do not include COHERENT in our analysis. We remark, however, that near-term expansions to the neutrino program at the Spallation Neutron Source could yield meaningful new constraints in this parameter space \cite{Akimov:2018ghi}.

Lastly, we consider the CONUS experiment \cite{CONUStalk}, for which an official exclusion result does not yet exist. We estimate the sensitivity of the CONUS experiment to oscillations involving a sterile neutrino following the procedures employed in Refs.~\cite{Farzan:2018gtr,Berryman:2018jxt}; we present this analysis fully in Appendix \ref{app:CONUS}, but provide some relevant details here. We consider two benchmark configurations for CONUS. The first is the nominal CONUS configuration, consisting of 4 kg of natural germanium and a recoil threshold of 1.2 keV taking data over one year; we call this configuration ``CONUS.'' The second is a more aggressive configuration, consisting of 100 kg of 88\% enriched germanium with a threshold of 0.1 keV, taking data for five years; we call this configuration ``CONUS100.'' We also assume that systematic uncertainties will improve from $\mathcal{O}(1\%)$ down to $\mathcal{O}(0.1\%)$. Other relevant details are summarized in App.~\ref{app:CONUS}. 

The resulting sensitivities for the CONUS and CONUS100 scenarios are shown in long-dashed blue and short-dashed light blue, respectively, in Fig.~\ref{fig:CENNS}. The sensitivity of the default CONUS configuration is relatively weak --- it is comparable to existing bounds from $\overline{\nu}_ee$ scattering. The CONUS collaboration has completed their data taking and is expected to be releasing their results in the near future, at which point it will be possible to assess the quality of the simulation performed here. However, it seems unlikely that the CONUS experiment will be able to probe the best-fit region to current short-baseline oscillation data \cite{Dentler:2018sju}, represented by a black star in Fig.~\ref{fig:CENNS}.

The projection for CONUS100, however, is much more optimistic. Our analysis suggests that this experiment would be able to conclusively probe the parameter space preferred by short-baseline anomalies and would be competitive with other next-generation CE$\nu$NS proposals. We remind the reader that this configuration is intended to be more optimistic than could likely be achieved, in order to demonstrate the extent to which these kinds of experiments could ultimately probe the cosmology of a sterile neutrino.


\section{Cosmology with Sterile Neutrinos}
\label{sec:cosmology}
\setcounter{equation}{0}

We outline the formalism for studying neutrino oscillations in the early Universe in order to keep this manuscript reasonably self-contained. For a more detailed account, see, for example, Refs.~\cite{Stodolsky:1986dx,Enqvist:1991qj,Raffelt:1992uj,McKellar:1992ja,9781107013957,Hannestad:2012ky,Volpe:2013jgr,Hannestad:2013wwj}.

In the early Universe, neutrino interactions with the matter background and with other neutrinos render the description of neutrino oscillations in the Sec.~\ref{sec:formalism} insufficient. Instead, one must solve for the evolution of the density matrix $\rho_{\vec{p}}$ ($\overline{\rho}_{\vec{p}}$) that describes an ensemble of neutrino (antineutrino) states with momentum $\vec{p}$. The equation of motion for $\rho_{\vec{p}}$ is given by \cite{Raffelt:1992uj,McKellar:1992ja,9781107013957}
\begin{equation}
i \frac{d\rho_{\vec{p}}}{dt} = \left[ \Omega^0_{\vec{p}}, \rho_{\vec{p}} \right] + \left[ \Omega^{\text{int}}_{\vec{p}}, \rho_{\vec{p}} \right] + \bold{C}\left[\rho_{\vec{p}}, \overline{\rho}_{\vec{p}}  \right];
\end{equation}
a similar equation exists for $\overline{\rho}_{\vec{p}}$. The $\Omega^0_{\vec{p}}$ term describes vacuum oscillations, while $\Omega^\text{int}_{\vec{p}}$ describes the matter potential, including contributions from other neutrinos. The collision term $\bold{C}\left[ \rho_{\vec{p}}, \overline{\rho}_{\vec{p}} \right] $ describes incoherent, inelastic interactions of neutrinos with their environment and with each other; it contains integrals involving $\rho_{\vec{p}}$ and $\overline{\rho}_{\vec{p}}$, and is the source of much of the technical difficulty in solving the evolution of the (anti)neutrino fluid.

To simplify the problem, we consider mixing between only two neutrinos, which we call active, $a$, and sterile, $s$,
\begin{equation}
\left( \begin{array}{c} \nu_a \\ \nu_s \end{array} \right) = \left( \begin{array}{cc} \cos \theta & -\sin \theta \\ \sin \theta & \cos \theta \end{array} \right) \left( \begin{array}{c} \nu_1 \\ \nu_2 \end{array} \right),
\end{equation}
where $\theta$ is the active-sterile mixing angle and $\nu_{1}$ and $\nu_{2}$ are neutrino mass eigenstates with masses $m_{1}$ and $m_{2} > m_1$, respectively. It is for this reason that we only consider one of the $\phi_{i4} \, (i = 1, \, 2)$ to be nonzero, and we identify $\theta = \phi_{i4}$ above, depending on which active species is being considered. Suppressing the subscript $\vec{p}$, the two-by-two matrix $\rho$ can be decomposed as
\begin{equation}
\rho = \frac{1}{2} f_0 \left( P_0 + \vec{\sigma} \cdot \vec{P} \right),
\end{equation}
where $\vec{\sigma}$ is the vector of the Pauli matrices and $f_0$ is the Fermi-Dirac distribution with vanishing chemical potential. The evolution of $P_0$ and the Bloch vector $\vec{P}$ are governed by
\begin{align}
\frac{d P_0}{dt} = & \, \, R^{(a)}, \\
\frac{d \vec{P}}{dt} = & \, \, \left(\vec{B} + \vec{V}^{(a)} \right) \times \vec{P} - D^{(a)} \left( P_x \, \hat{x} + P_y \, \hat{y} \right) \nonumber \\
& + R^{(a)} \, \hat{z}, 
\end{align}
where $\vec{B} \equiv \left(\frac{\Delta m^2}{2 p}\right) \left(\sin 2\theta, \, 0, \, - \cos 2\theta \right)$ and $\Delta m^2 \equiv m_2^2 - m_1^2$. A similar decomposition exists for $\overline{\rho}$, into $\overline{P}_0$ and $\vec{\overline{P}}$.

The potential $\vec{V}^{(a)}$ depends on whether $\nu_a$ is electron-type or muon-/tau-type, as $\nu_e$ has charged-current interactions with a background of electrons that $\nu_\mu$ and $\nu_\tau$ do not. The matter potential is \cite{Hannestad:2012ky,Hannestad:2013wwj}
\begin{align}
\vec{V}^{(a)} & = \left( V^{(a)}_1 + V^{(a)}_L \right) \hat{z}, \\
V^{(a)}_1 & = -\dfrac{7 \pi^2 G_F}{45\sqrt{2} M_Z^2} p \, T^4 \left( n_{\nu_a} + n_{\overline{\nu}_a} \right) g_a, \\
V^{(a)}_L & = \frac{2\sqrt{2} \zeta(3)}{\pi^2} G_F T^3 L^{(a)},
\end{align}
where $n_f$ is the number density of species $f$. The constant $g_a$ is either $g_{\mu,\tau} = 1$ or $g_e = 1 + 4\sec^2 \theta_W/ \left( n_{\nu_e} + n_{\overline{\nu}_e} \right)$, and the lepton asymmetries $L^{(a)}$ are given by
\begin{align}
\label{eq:L_elec}
L^{(e)} = & \, \,  \left( \frac{1}{2} + 2 \sin^2 \theta_W \right) L_e +  \left( \frac{1}{2} - 2 \sin^2 \theta_W \right) L_p \nonumber \\
& - \frac{1}{2} L_n + 2 L_{\nu_e} + L_{\nu_\mu} + L_{\nu_\tau}, \\
L^{(\mu,\tau)} = & \,\, L^{(e)} - L_e - L_{\nu_e} + L_{\nu_\mu, \nu_\tau},
\label{eq:L_mutau}
\end{align}
with $L_f \equiv (n_f - n_{\overline{f}}) n^{\text{eq}}_f/n^{\text{eq}}_\gamma$, where $n^{\text{eq}}_f$ is the equilibrium number density of $f$. The damping function $D^{(a)}$ characterizes the loss of quantum coherence from interactions with the background, and, assuming thermal equilibrium, is approximately given by
\begin{equation}
D^{(a)} \approx \frac{1}{2} \Gamma^{(a)}, \qquad \text{where } \Gamma^{(a)} = C^{(a)} G_F^2 T^4 p;
\end{equation}
$C^{(e)} \approx 1.27$ is used for $\nu_e$ and $C^{(\mu, \tau)} \approx 0.92$ is used with $\nu_{\mu, \tau}$ \cite{Enqvist:1991qj}. Lastly, the repopulation function $R^{(a)}$ is given approximately by
\begin{equation}
R^{(a)} \approx \Gamma^{(a)} \left[ \frac{f_{\text{eq}}(p, \mu_{\nu_a})}{f_0} - \frac{1}{2} \left( P_0 + P_z \right) \right],
\end{equation}
where $f_{\text{eq}}(p, \mu_{\nu_a})$ is the equilibrium Fermi-Dirac distribution with chemical potential $\mu_{\nu_a}$.

It is convenient for numerical evaluation to define the quantities
\begin{align}
P^\pm_i & = P_i \pm \overline{P}_i, \\
P^\pm_a & = P^\pm_0 + P^\pm_z, \\
P^\pm_s  & = P^\pm_0 - P^\pm_z,
\end{align}
for $i = 0, \, x, \, y, \, z$. Their evolution is given by
\begin{align}
\label{eq:EOM1}
\frac{d P^\pm_a}{dt} = & \,\, B_x P^\pm_y + \Gamma_a \left( 2 f_{\text{eq}}^\pm/f_0 - P_a^\pm \right), \\
\frac{d P_s^\pm}{dt} = & \,\, -B_x P_y^\pm, \\
\frac{d P_x^\pm}{dt} = & \,\, -(B_z + V^{(a)}_1) P_y^\pm - V_L^{(a)} P_y^\mp - D^{(a)} P_x^\pm, \\
\frac{d P_y^\pm}{dt} = & \,\, (B_z +V^{(a)}_1) P_x^\pm + V_L^{(a)} P_x^\mp \nonumber \\
& - \frac{1}{2} B_x \left(P_a^\pm - P_s^\pm \right) - D^{(a)} P_y^\pm, \label{eq:EOM4}
\end{align}
where $f_{\text{eq}}^\pm \equiv f_{\text{eq}}(p, \mu_{\nu_a}) \pm f_{\text{eq}}(p, -\mu_{\nu_a})$.

\begin{figure}[!tbp]
\includegraphics[width=\linewidth]{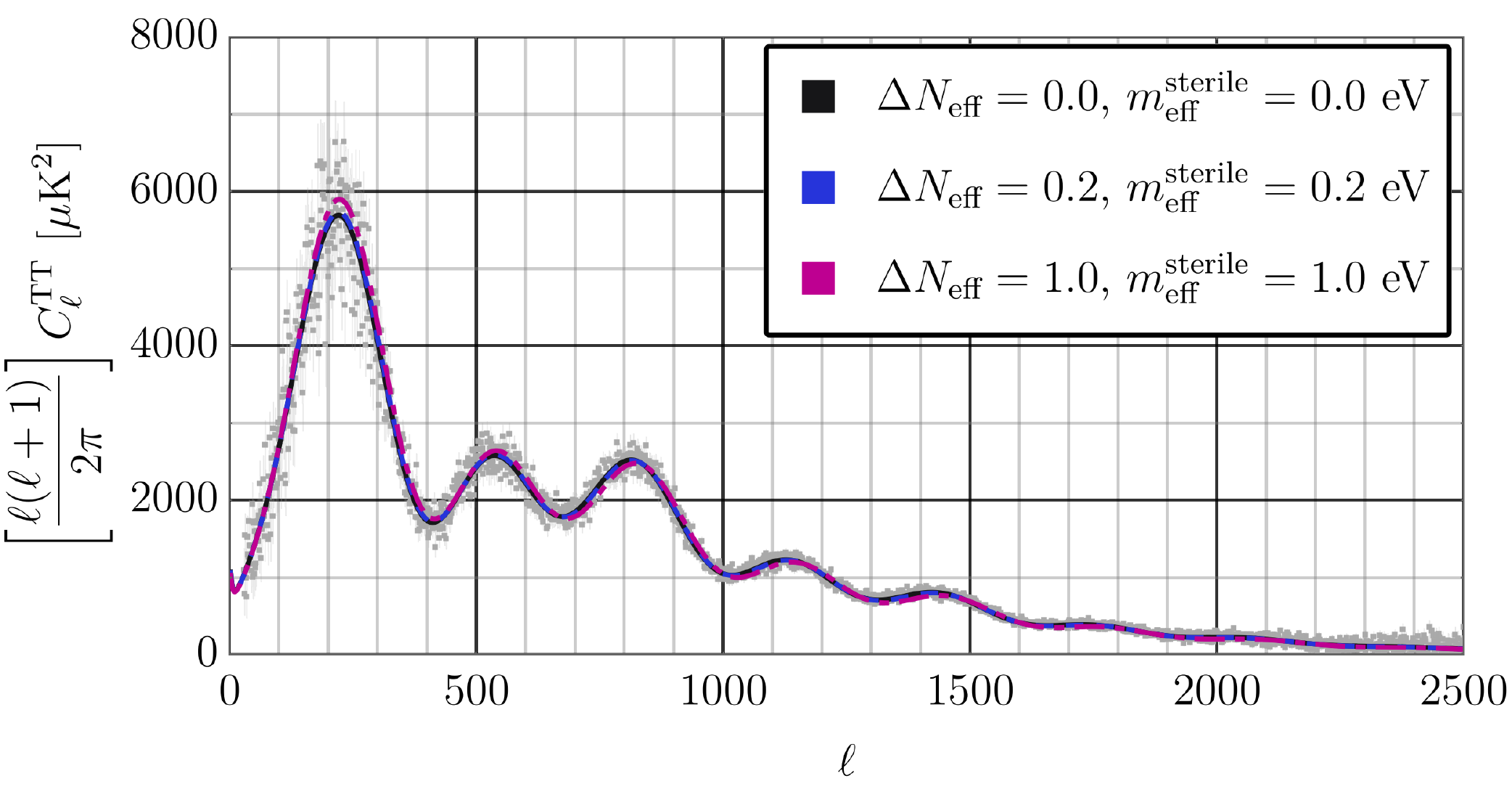}
\caption{The CMB TT power spectrum for the base $\Lambda$CDM cosmology and for two benchmark sterile neutrino scenarios. Curves produced using the \textsc{camb} module \cite{Lewis:1999bs}. The gray squares represent the Planck 2015 measurements and their uncertainties \cite{Ade:2015xua}.}
\label{fig:CMB_Cs}
\includegraphics[width=\linewidth]{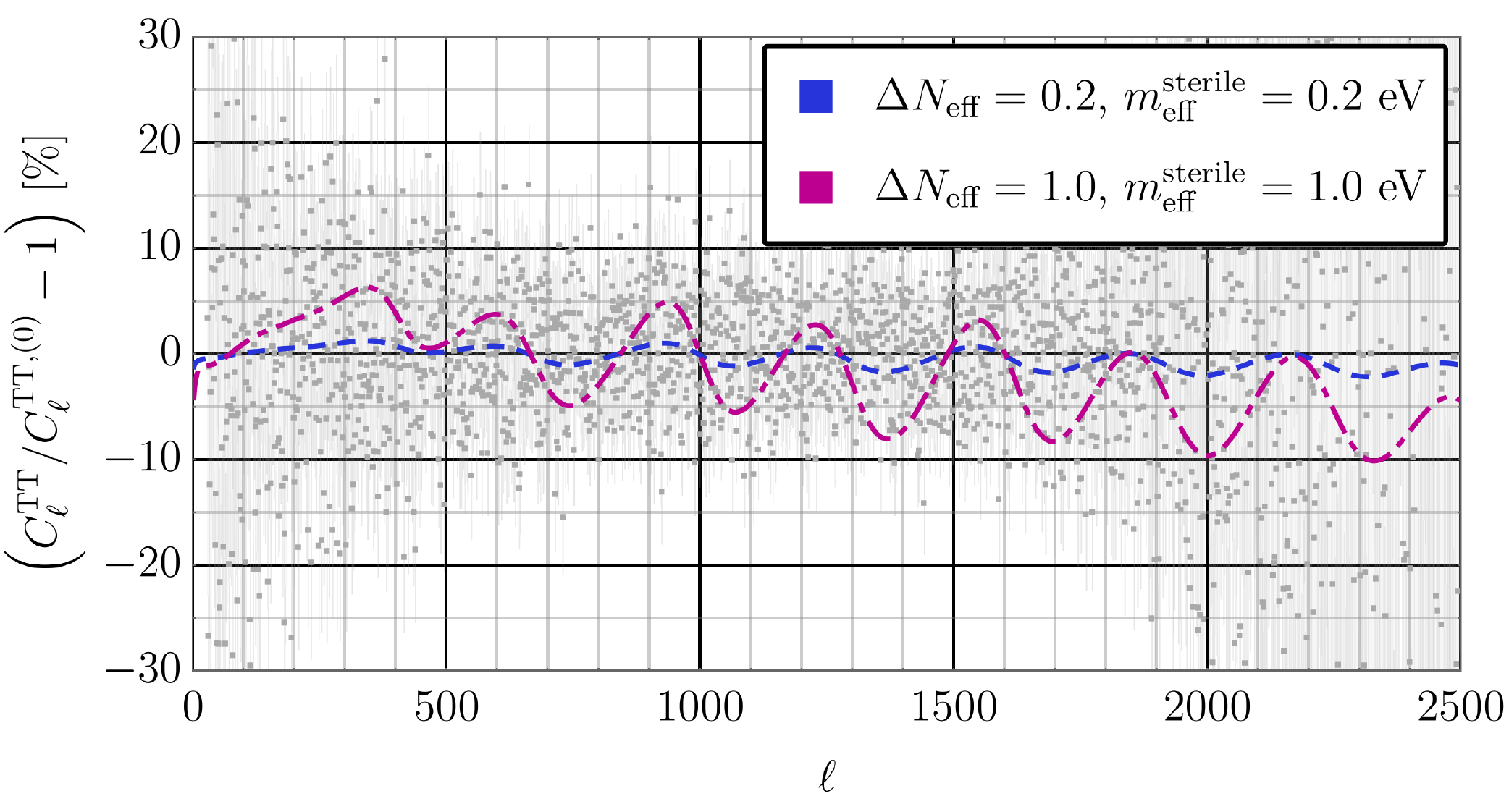}
\caption{The percent differences between our benchmark sterile neutrino scenarios in Fig.~\ref{fig:CMB_Cs} relative to the base $\Lambda$CDM cosmology. The gray squares represent the Planck 2015 measurements and their uncertainties \cite{Ade:2015xua}.}
\label{fig:CMB_diffs}
\end{figure}

The standard $\Lambda$CDM cosmology can be augmented with a sterile neutrino through the introduction of two parameters: $\Delta N_{\rm eff} \equiv N_{\rm eff} - 3.046$ and $m^{\rm eff}_{\rm sterile} \equiv (94.1 \text{ eV}) \Omega_{\rm sterile } h^2$. In Fig.~\ref{fig:CMB_Cs}, we show the CMB TT power spectrum for the $\Lambda$CDM cosmology, as well as for two benchmark values of $\Delta N_{\rm eff}$ and $m^{\rm eff}_{\rm sterile}$; the best-fit points for the six parameters of the base $\Lambda$CDM cosmology are shown in Table \ref{table:cosmology}. In Fig.~\ref{fig:CMB_diffs}, we show the percent deviation of the two latter scenarios relative to the former. In both figures, the gray squares represent the Planck 2015 measurements of the temperature spectrum and their uncertainties \cite{Ade:2015xua}.

\begin{table}[!t]
\begin{center}
\begin{tabular}{|c||c|}\hline
Parameter & Value \\ \hline \hline
$\Omega_b h^2$ & $0.02225$ \\ \hline
$\Omega_c h^2$ & $0.1198$ \\ \hline
$100 \theta_{\rm MC}$ & $1.04077$ \\ \hline
$\tau$ & $0.079$ \\ \hline
$\ln(10^{10} A_s)$ & $3.094$ \\ \hline
$n_s$ & $0.9645$ \\ \hline
\end{tabular}
\caption{Best-fit parameters to the base $\Lambda$CDM cosmology used to make Figs.~\ref{fig:CMB_Cs} and \ref{fig:CMB_diffs}, from the ``TT,TE,EE+lowP'' fit to Planck data \cite{Ade:2015xua}.}
\label{table:cosmology}
\end{center}
\end{table}

One can calculate $\Delta N_{\rm eff}$ at any point in the evolution of the Universe via
\begin{equation}
\label{eq:Neff}
\Delta N_{\rm eff} = \frac{\int dx \, x^3 f_{\text{eq}}(x, \mu = 0) \, P_s^+(x)}{4 \int dx \, x^3 f_{\text{eq}}(x, \mu = 0)},
\end{equation}
where $x \equiv p/T$. Ref.~\cite{Bridle:2016isd} discusses two ways in which the mass of the fourth neutrino, $m_4$, can be related to $m^{\rm eff}_{\rm sterile}$. The first is the Dodelson-Widrow mechanism \cite{Dodelson:1993je}, wherein the two are related via $m^{\rm eff}_{\rm sterile} = \Delta N_{\rm eff} \, m_4$. In this work, however, we will instead follow Ref.~\cite{Bridle:2016isd} and use the relation
\begin{equation}
\label{eq:meff}
m^{\rm eff}_{\rm sterile} = (\Delta N_{\rm eff})^{3/4} \, m_4,
\end{equation}
which assumes that sterile neutrinos are produced thermally in the early Universe. We have verified that our results do not depend strongly on this assumption.

\begin{figure}[!t]
\includegraphics[width=\linewidth]{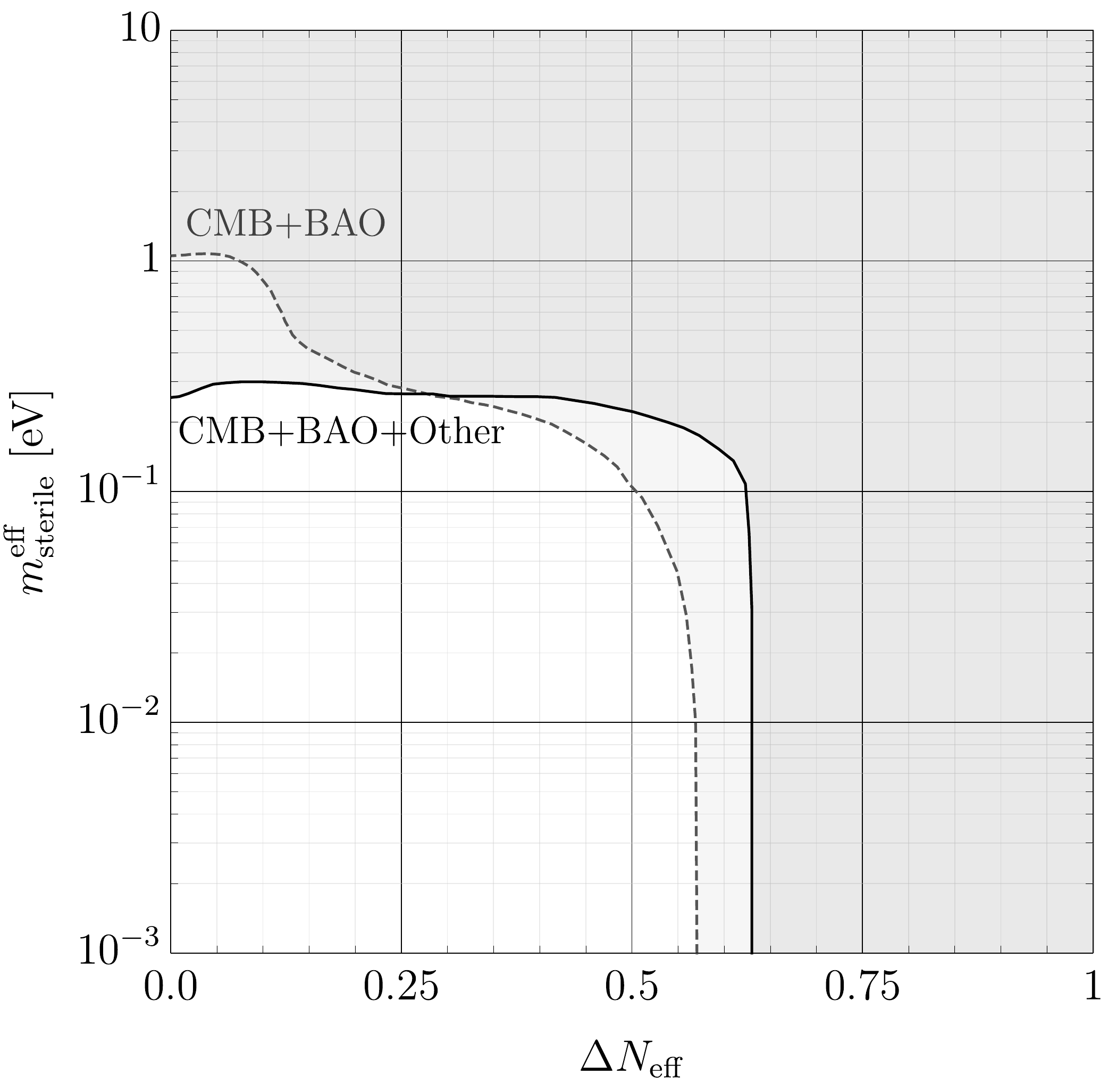}
\caption{The 2$\sigma$ C.L. exclusions in the \NMplane plane derived in Ref.~\cite{Feng:2017nss} for the CMB+BAO (dotted gray) and CMB+BAO+Other (black) datasets; see text for details.}
\label{fig:planck_data}
\end{figure}

We consider two bounds in the \NMplane parameter space from cosmological data, calculated and presented in Ref.~\cite{Feng:2017nss}. The first is a combined analysis of the Planck 2015 CMB temperature and polarization (TT, TE and EE) spectra \cite{Ade:2015xua}, for both high and low $\ell$, and baryon acoustic oscillation (BAO) data \cite{Beutler:2011hx,Ross:2014qpa,Cuesta:2015mqa}. Following Ref.~\cite{Feng:2017nss}, we call this ``CMB+BAO," shown in dotted gray in Fig.~\ref{fig:planck_data}. This limit is similar to, but stronger than, a similar limit derived in Ref.~\cite{Bridle:2016isd}. The second adds to this the value of the Hubble constant measured in Ref.~\cite{Riess:2016jrr} as a prior, as well as Planck cluster \cite{Ade:2015fva} and lensing data \cite{Ade:2015zua} and weak lensing data from Ref.~\cite{Heymans:2013fya}, which we call ``CMB+BAO+Other." This is shown in black in Fig.~\ref{fig:planck_data}.

The shift in the astrophysical/cosmological limit with the inclusion of the ``Other'' data sets stems, in part, from tension in measurements of the Hubble constant; Planck measures a smaller value \cite{Ade:2015xua} than local measurements indicate \cite{Riess:2016jrr}. The origin of this discrepancy is as of yet unclear, but its effect is to allow for a modestly larger value of $\Delta N_{\rm eff}$ at the expense of a more stringent limit on $m_{\rm sterile}^{\rm eff}$. This is to be expected from, for instance, Figs.~29-31 in Ref.~\cite{Ade:2015xua} -- larger values of the Hubble constant result in a preference for larger $\Delta N_{\rm eff}$. 

While the most recent Planck data release \cite{Aghanim:2018eyx} assuredly implies a more stringent constraint in this space, we do not consider it here. In the absence of the Planck 2018 likelihood function, it is not possible to quantitatively determine how updated measurements of the local Hubble constant \cite{riess2,Riess:2018byc} will shift the limit in the \NMplane plane. We note, however, that the discrepancy between local and cosmological determinations of the Hubble constant persists in these recent measurements (see Ref.~\cite{Bernal:2016gxb}); the resolution of this puzzle has important implications for constraints on additional neutrinos \cite{Guo:2018ans,Carneiro:2018xwq}.

We assume throughout that the lightest neutrino mass vanishes; for a normal neutrino mass ordering, this means $m_1 = 0$. Furthermore, we have $m_2 = \sqrt{\Delta m_{21}^2} \approx 0$, $m_3 = \sqrt{\Delta m_{31}^2}$ and $m_4 = \sqrt{\Delta m_{41}^2}$. The Planck analysis assumes that any excess in neutrino mass is attributable to a single additional state -- in our case, a sterile neutrino. If $m_1 > 0$ or if neutrino masses were arranged in an inverted hierarchy, then the constraints in Fig.~\ref{fig:planck_data} strengthen. 

The primary objective of this work is to translate limits that oscillation experiments place in the $\sin^2 2\theta_{\alpha \alpha}$ -- $\Delta m^2_{41}$ plane ($\alpha = e, \, \mu$) into limits in the \NMplane plane. We do so using the code \texttt{LASAGNA} \cite{Hannestad:2013wwj} to solve for the evolution of the neutrino fluid in the early Universe. We use as inputs the points in the $\sin^2 2\theta_{\alpha\alpha}$--$\Delta m^2_{41}$ plane that comprise the experimental exclusions and sensitivities. For each point, \texttt{LASAGNA} calculates the evolution of the Bloch vector from $T$ = 40 MeV to $T$ = 1 MeV -- roughly the temperature of the Universe when neutrinos decouple from the Standard Model bath -- using the equations of motion in Eqs.~\eqref{eq:EOM1}--\eqref{eq:EOM4}. We determine $\Delta N_{\rm eff}$ via Eq.~\eqref{eq:Neff} and use this to calculate $m^{\rm eff}_{\rm sterile}$ using Eq.~\eqref{eq:meff}. We assume throughout that the initial lepton asymmetry of the Universe is zero.

We are careful to distinguish between experiments that bound $\sin^2 2\theta_{\mu\mu}$ and those that bound $\sin^2 2\theta_{ee}$ because of differences in $\nu_e$ -- $\nu_s$ oscillations and $\nu_\mu$ -- $\nu_s$ oscillations in the early Universe. The effective lepton asymmetry electron neutrinos experience is different than for muon neutrinos -- see Eqs.~\eqref{eq:L_elec} and \eqref{eq:L_mutau} -- resulting in a different matter potential. We consider bounds on $\sin^2 2\theta_{\mu\mu}$ from accelerator neutrino experiments and bounds on $\sin^2 2\theta_{ee}$ coming from reactor and other low-threshold neutrino experiments; the cosmological parameter space is the same in each case, but we underscore that these constitute distinct hypotheses.


\section{Results}
\label{sec:results}
\setcounter{equation}{0}

\subsection*{Accelerator Neutrino Experiments}

\begin{figure}[!t]
\includegraphics[width=\linewidth]{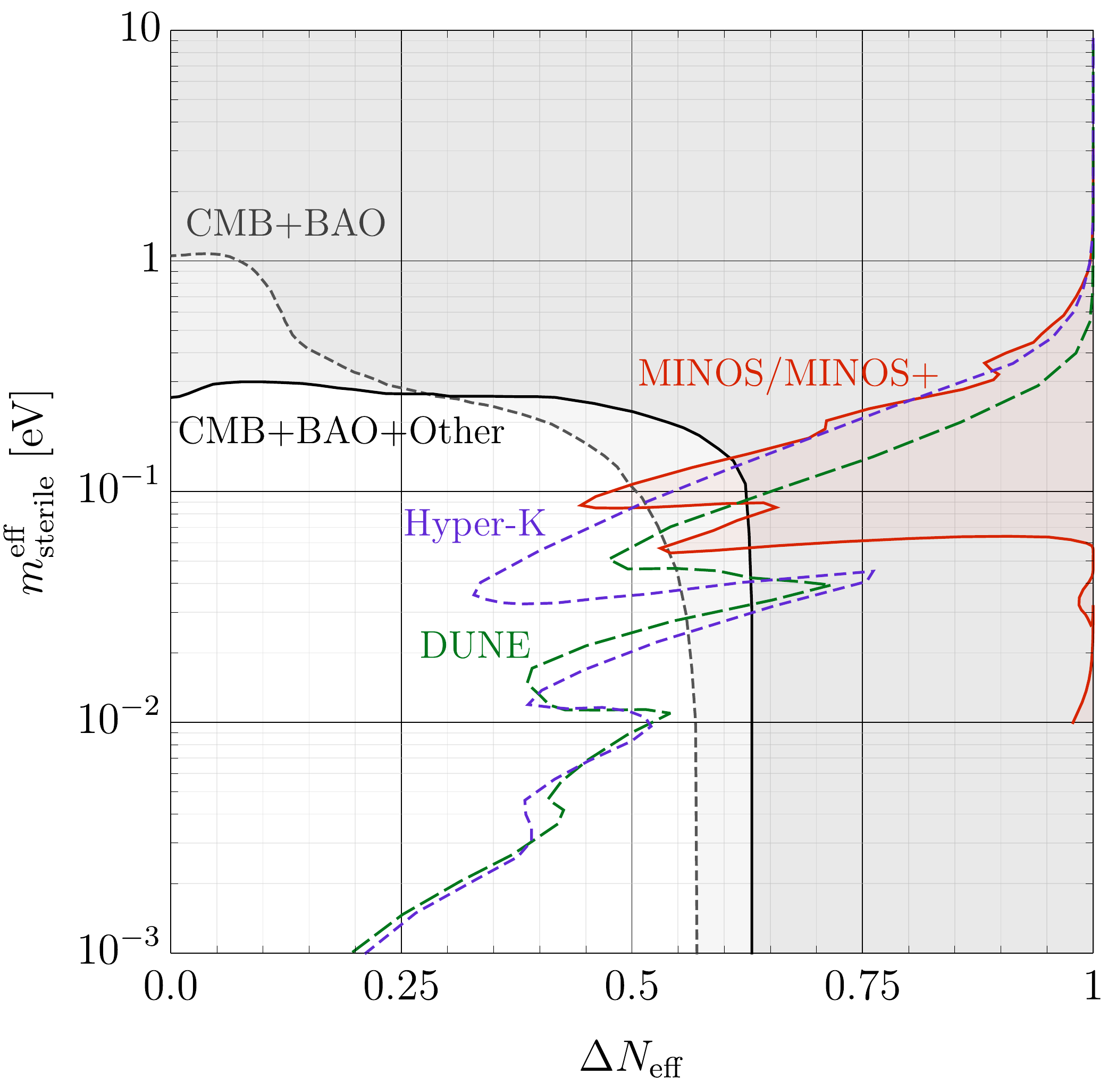}
\caption{Constraints in the \NMplane plane derived from accelerator experiments: MINOS/MINOS+ (red, 90\% C.L.), DUNE (long-dashed green, 95\% C.L.) and Hyper-K (short-dashed purple, 95\% C.L.).}
\label{fig:accelerator_results}
\end{figure}

In Figure~\ref{fig:accelerator_results}, we show the resulting exclusions/sensitivities in the \NMplane plane for accelerator neutrino experiments. Evidently, MINOS/MINOS+ can probe new regions of the cosmological parameter space -- particularly compared to the CMB+BAO+Other analysis -- but the excursion into untouched parameter space is not quite as strong as previously reported in Ref.~\cite{Bridle:2016isd}. There, a bound in the \NMplane plane from MINOS was derived from the analysis of Ref.~\cite{Huang:2015qia}. This analysis, however, takes some oscillation parameters to be fixed, including the solar parameters ($\Delta m_{21}^2$ and $\theta_{12}$), the reactor angle ($\theta_{13}$) and the $CP$-odd phase $\delta_{CP}$. This bound is overly optimistic; the uncertainties on the fixed parameters are nontrivial, and ignoring them artificially enhances the confidence of the final result. Therefore, we advocate that the red curve in Fig.~\ref{fig:accelerator_data} is a more accurate representation of the capabilities of MINOS/MINOS+ as a probe of neutrino cosmology.

Also shown in Fig.~\ref{fig:accelerator_results} are our projected 95\% C.L. sensitivity limits for DUNE and Hyper-K. These long-baseline experiments have comparable sensitivity to the $m_{\rm sterile}^{\rm eff} \lesssim 10^{-2}$ eV portion of the \NMplane plane that is unconstrained by the CMB+BAO and CMB+BAO+Other analyses, while Hyper-K has slightly more sensitivity than DUNE in the region $m_{\rm sterile}^{\rm eff} \sim 10^{-1}$ eV. The upshot is that these experiments (will) have a capability to cut into a modest portion of the cosmological parameter space that is not currently probed by Planck and other astrophysical/cosmological experiments. Several comments are in order:
\begin{enumerate}
\item These experiments are able to probe new parts of this parameter space primarily because of their sensitivities to oscillations with $\Delta m_{41}^2 \sim 10^{-2}-10^{-1}$ eV$^{2}$ and $\sin^2 2\theta_{\mu\mu} \sim {\rm few} \times 10^{-2}$. Moreover, the modest sensitivity that DUNE and Hyper-K will have in the $\Delta m_{41}^2 \sim 10^{-4}-10^{-3}$ eV$^{2}$ regime translates into a substantial sensitivity to $m_{\rm sterile}^{\rm eff} \lesssim \mathcal{O}(10^{-1})$ eV$^2$.
\item The inclusion of ``Other" data with the CMB+BAO data set relaxes the bound on $\Delta N_{\rm eff}$. However, terrestrial oscillation experiments (will) provide some coverage in the region in which the cosmological constraints are relaxed.
\item While the improved sensitivity is interesting, DUNE and Hyper-K will not start collecting data until at least the late 2020s. In the interim, next-generation projects like CMB-S4 \cite{Abazajian:2016yjj} will continue to whittle away at the available parameter space; CMB-S4 is expected to be completed by the mid 2020s. Even if the latter is delayed, by the time DUNE and Hyper-K have data to analyze, they will not be able to substantively probe new parameter space. On the other hand, should CMB-S4 be able to exclude, say, nonzero $N_{\rm eff}$ at high confidence, then these terrestrial experiments may be able to independently probe whether or not the relic is a sterile neutrino.
\end{enumerate}

\subsection*{Reactor Antineutrino Experiments}

\begin{figure}[!t]
\includegraphics[width=\linewidth]{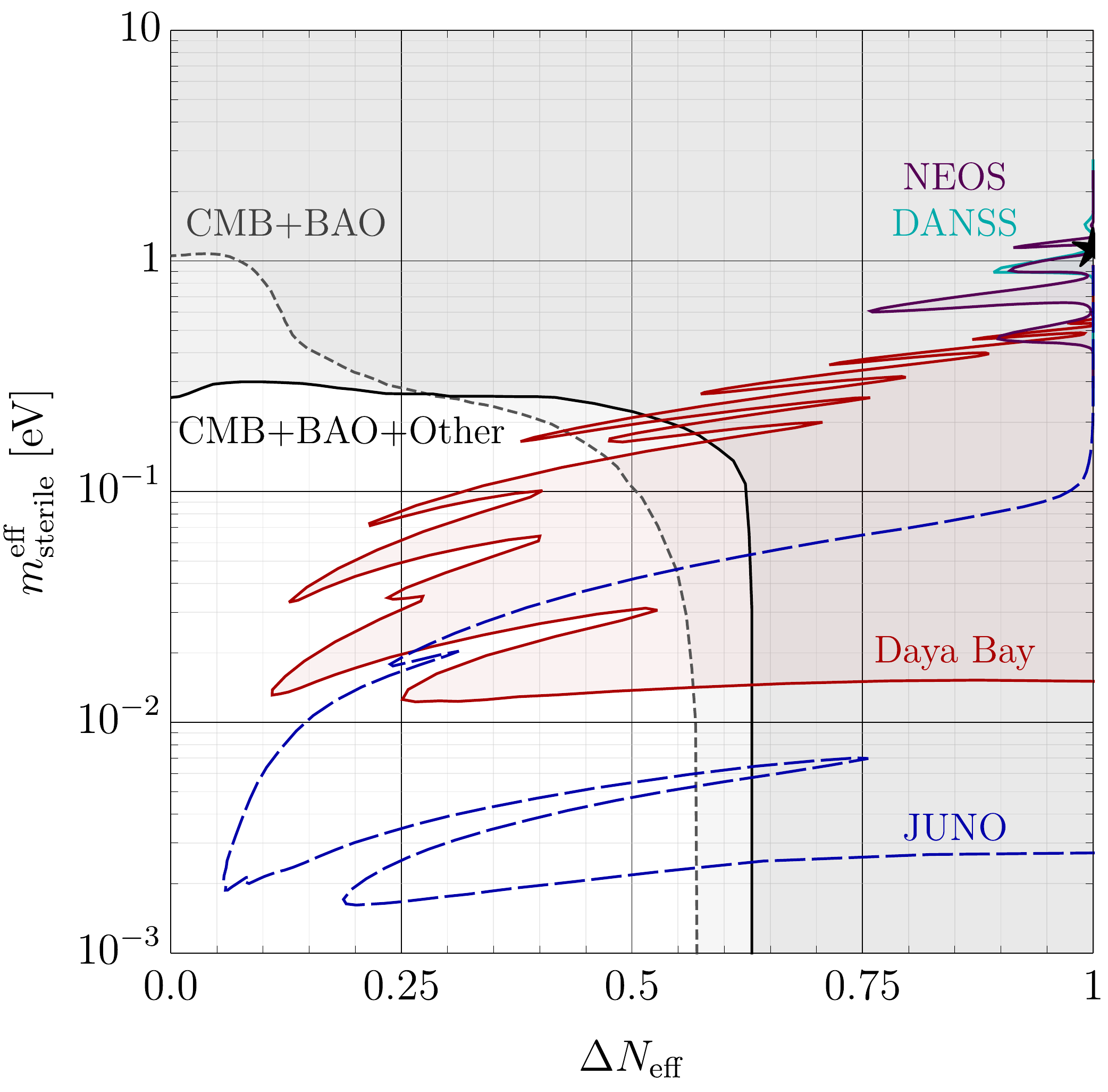}
\caption{Constraints in the \NMplane plane derived from reactor experiments: DANSS (cyan, 95\% C.L.), Daya Bay (red, 95\% C.L.), NEOS (purple, 90\% C.L.) and JUNO (long-dashed blue, 95\% C.L.). The black, five-pointed start represents the best-fit point from Ref.~\cite{Dentler:2018sju}.}
\label{fig:reactor_results}
\end{figure}

Limits in the \NMplane plane from reactor antineutrino experiments are shown in Figure~\ref{fig:reactor_results}, from which we deduce several important features. The first is that Daya Bay is (and JUNO will be) able to probe parts of this space not currently constrained by either cosmological data set. However, while DUNE, Hyper-K and JUNO are still years away from taking data, \emph{Daya Bay can already rule out a significant portion of the \NMplane plane to which astrophysical/cosmological data are currently insensitive.} This is one of the key conclusions of this work. On the basis of Fig.~37 in Ref.~\cite{Aghanim:2018eyx}, it is likely that this persists in the most recent data release from Planck, especially given that Daya Bay, too, possesses an updated data set \cite{Adey:2018zwh,Adey:2019ywk}. This is consistent with findings in Ref.~\cite{Knee:2018rvj}, where a different set of astrophysical/cosmological constraints were considered.

The second feature is that the bounds from DANSS and NEOS are scarcely visible in this plot; they're compressed against the boundary at $\Delta N_{\rm eff}=1$, excluded at high significance by both the CMB+BAO and CMB+BAO+Other constraints. These experiments dominate the fit to reactor data that results in the best-fit point shown, again, by the black, five-pointed star in Fig.~\ref{fig:reactor_results}. Taking this at face value implies that astrophysics and cosmology already rule out a sterile neutrino with $\Delta m_{41}^2 \sim \mathcal{O}(1)$ eV$^2$.

\subsection*{Low-Threshold Experiments}

\begin{figure}[!t]
\includegraphics[width=\linewidth]{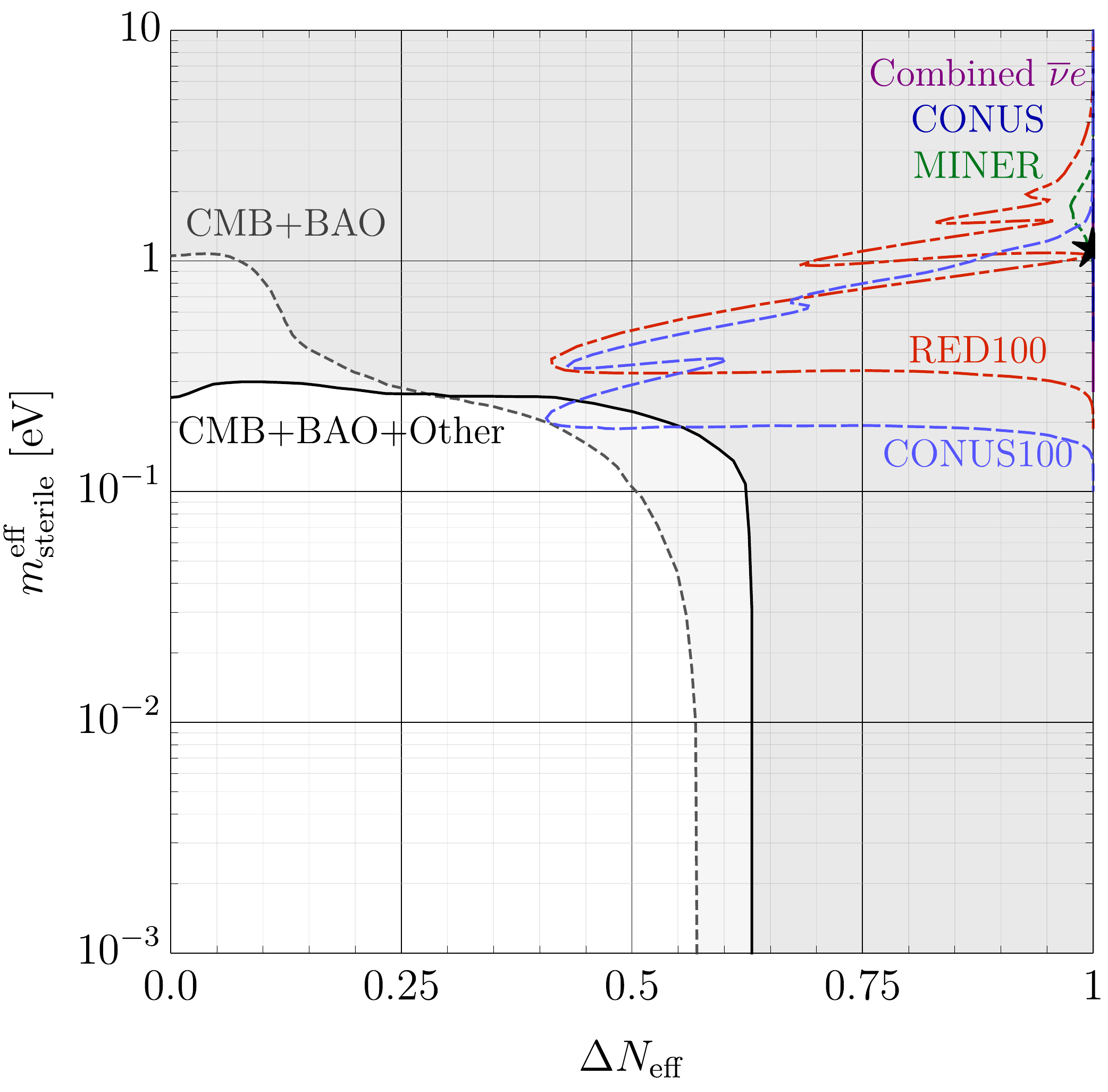}
\caption{Constraints in the \NMplane plane derived from low-threshold experiments: $\overline{\nu}e$ scattering (purple, 90\% C.L.), RED100 (dot-dashed red, 90\% C.L.), MINER (double-dot-dashed green, 90\% C.L.), CONUS (long-dashed blue, 95\% C.L.) and CONUS100 (short-dashed light blue, 95\% C.L.). The black, five-pointed start represents the best-fit point (BFP) from the global analysis of Ref.~\cite{Dentler:2018sju}.}
\label{fig:CEvNS_results}
\end{figure}

Figure \ref{fig:CEvNS_results} depicts the limits from the low-threshold experiments we have considered. The sensitivity of MINER is scarcely visible on this plot; the sensitivity of RED100 is visible, but improves on neither the CMB+BAO nor the CMB+BAO+Other exclusions. Recall that we selected these specific incarnations of the RED100 and MINER experiments because they were the most optimistic proposals for probing the $\sin^2 2 \theta_{ee}$--$\Delta m_{41}^2$ parameter space. However, on the basis of Figure \ref{fig:CEvNS_results}, it seems that these experiments will not offer a fresh perspective on cosmology.

Also included in Fig.~\ref{fig:CEvNS_results} are the exclusion from $\overline{\nu}e$ scattering and the sensitivity of CONUS. However, neither is visible on this plot; these constraints deviate from $\Delta N_{\rm eff} = 1$ by at most $\sim {\rm few} \times 10^{-5}$. On the other hand, CONUS100 is capable of reaching a sliver of the parameter space to which the CMB+BAO+Other constraint is insensitive. However, this configuration has been chosen to be extremely optimistic -- arguably unrealistically so -- and even under these conditions, it is barely able to improve limits on the cosmology of a sterile neutrino. Moreover, the most recent Planck data and CMB-S4 would likely render these experiments totally impotent when it comes to excluding cosmological parameter space. 

These results and those of the previous subsection present a problem. As we have mentioned, the best-fit point to the reactor antineutrino anomaly -- the black star in Figs.~\ref{fig:reactor_results} and \ref{fig:CEvNS_results} -- is strongly disfavored by the CMB+BAO(+Others) data sets. Indeed, we have seen how experiments that either provide evidence for the anomaly (DANSS, NEOS, etc.) or are designed to probe the anomaly (RED100, MINER, CONUS100, etc.) are powerless to challenge the astrophysical/cosmological experiments in this parameter space. We are then left to question if (and how) these data sets can be rendered consistent. We enumerate a few possible resolutions.
\begin{enumerate}
\item The reactor antineutrino anomaly is an aberration. Determining the reactor antineutrino flux is a complicated business; there is no shortage of ways in which theoretical calculations and experimental measurements could go awry. See, for instance, Refs.~\cite{Mueller:2011nm,Huber:2011wv,Hayes:2015yka,Huber:2016fkt,Hayes:2016qnu,Giunti:2016elf,Huber:2016xis} for more details. We note that, of the reactor experiments we have considered, DANSS, Daya Bay and NEOS\footnote{Note that NEOS uses the Daya Bay spectrum to normalize their antineutrino spectrum. The different effective fuel fractions at these experiments introduces a small amount of dependence on the flux model.} employ a near detector to reduce flux-related systematics; the rest, however, depend on theoretical predictions of the flux. 
\item Our understanding of cosmology is incomplete. This seems unlikely, but it is not altogether impossible that something dramatic could have happened in the early Universe that the standard $\Lambda$CDM cosmology doesn't capture.
\item The framework we have employed is insufficient. The two-neutrino approximation is useful to solve for the evolution of the neutrino fluid, but it may be missing some important physics. A more complete calculation would be more intensive, but potentially worth the effort.
\item Neutrinos have extra interactions -- affecting either the active or sterile flavors -- that may lead to a qualitatively different evolution of the neutrino fluid. This may be due to a new matter potential or because new degrees of freedom are relevant later in evolution of the Universe (see, for instance, Ref.~\cite{Cherry:2016jol,Archidiacono:2016kkh,Forastieri:2017oma,Song:2018zyl}).
\item The initial lepton asymmetry of the Universe, $L_0$, is large, $\sim \mathcal{O}(10^{-3}-10^{-2})$. This may suppress transitions to the sterile flavor, diminishing its contribution to $m_{\rm sterile}^{\rm eff}$ and $\Delta N_{\rm eff}$. The \texttt{LASAGNA} module is well suited to study the evolution of the (anti)neutrino fluid in the presence of an initial lepton asymmetry, and this has been studied in Ref.~\cite{Bridle:2016isd}; see also Ref.~\cite{Johns:2016enc}. We relegate a detailed study to future work.
\end{enumerate}

Whatever its cause, the observation of this tension is the other central conclusion of this work. \emph{There can be no satisfactory sterile neutrino solution to the reactor antineutrino anomaly that does not address this tension with astrophysical and cosmological measurements.}


\section{Conclusions}
\label{sec:conclusions}

We have considered the complementarity between terrestrial neutrino oscillation experiments -- specifically, accelerator and reactor experiments -- and cosmological experiments in probing the cosmological properties of a proposed sterile neutrino. Accelerator neutrino experiments have the potential to improve on current constraints, but given that these are years away from taking data, it is unlikely that they will improve on the constraints that will exist by then. Meanwhile, reactor experiments -- Daya Bay, specifically -- are already able to probe parameter space beyond the reach of cosmological experiments, a feature that may persist with updated measurements.

We have emphasized the tension between cosmological constraints and the mild preference for a sterile neutrino from reactor antineutrino experiments. This is highlighted by (1) the best-fit point from the global reactor analysis of Ref.~\cite{Dentler:2018sju} already being strongly disfavored by Planck, and (2) the inability of upcoming low-threshold neutrino scattering experiments to probe new parts of the cosmological parameter space. Several possibilities for resolving this tension have been enumerated, but we conclude that no solution to the reactor anomalies can be truly compelling if it does not address this tension with cosmology.

This analysis presented here is not, strictly speaking, entirely consistent. Some of the terrestrial experiments we have considered have been analyzed under the four-neutrino hypothesis (with only one active-sterile mixing angle allowed to be nonzero), whereas these bounds on sterile neutrino oscillation parameters have been translated into cosmological bounds using two-flavor oscillations. This could well be a fatal inconsistency -- it is logically possible that the reduction to two-flavor oscillations has oversimplified the system, so that crucial physics is being missed. In particular, this framework offers no opportunities to study the LSND \cite{Aguilar:2001ty} and MiniBooNE \cite{Aguilar-Arevalo:2013pmq} anomalies, since electron-neutrino appearance requires that two active-sterile mixing angles be nonzero. A more complete analysis is required, especially given the tension present in the sterile-neutrino interpretation of neutrino appearance and disappearance data \cite{Dentler:2018sju}.

The results of this work are meaningful heuristics, but that the hypotheses to which they apply -- that a sterile neutrino exists, but that only one possible active-sterile mixing angle is nonzero -- may be too simple to be physical. Moreover, the separation between electron-type and muon-type oscillations in our treatment of the neutrino fluid in the early Universe is awkward. A more comfortable arrangement would be to use bounds on the complete four-neutrino hypothesis from a variety of experiments simultaneously to derive bounds on sterile-neutrino cosmology. However, the primary limitation of this scheme is that solving for the evolution of three active and one sterile neutrino species (and their corresponding antineutrinos) in the early Universe is a technically daunting task.

This work is meant to illustrate the importance of interdisciplinary studies in neutrino physics. It is not entirely obvious, \emph{a priori}, that an experiment like Daya Bay can improve constraints in some part of the \NMplane plane, yet we have found this to be the case. In order to keep this work focused, we have considered only accelerator and reactor (anti)neutrino experiments and their interplay with astrophysics and cosmology. Even with the restriction that at most one active-sterile mixing angle is nonzero, similar analyses could be performed using (anti)neutrino disappearance results from solar \cite{Cleveland:1998nv,Aharmim:2005gt,Hosaka:2005um,Aharmim:2006kv,Cravens:2008aa,Aharmim:2008kc,Bellini:2008mr,Abdurashitov:2009tn,Kaether:2010ag,Abe:2010hy,Bellini:2011rx,Gando:2014wjd} or atmospheric \cite{Aartsen:2014yll,Wendell:2014dka,TheIceCube:2016oqi} experiments, as well as accelerator experiments that we have not considered \cite{Dydak:1983zq,Stockdale:1984cg,Cheng:2012yy,Adamson:2017zcg,Abe:2019fyx}. These bounds, however, are at best comparable to, and are generally weaker than, the limits we have presented.


\section*{Acknowledgements}
JMB thanks Andr\'e de Gouv\^ea, Patrick Huber and Kevin Kelly for useful discussions and acknowledges the support of the Colegio de F\'isica Fundamental e Interdisciplinaria de las Am\'ericas (COFI) Fellowship Program. JMB further thanks Kevin Kelly for providing the software for the simulations of DUNE and Hyper-K presented here. This work is supported by DOE Grant No.~de-sc0018327.


\appendix

\section{Simulations of DUNE and Hyper-Kamiokande}
\label{app:DUNE}

\begin{table}[!t]
\begin{center}
\begin{tabular}{|c||c|}\hline
Parameter & Value \\ \hline \hline
$\sin^2 \theta_{12}$ & $0.306$ \\ \hline
$\sin^2 \theta_{13}$ & $0.02166$ \\ \hline
$\sin^2 \theta_{23}$ & $0.441$ \\ \hline
$\delta_{CP}$ & $-\pi/2$ \\ \hline \hline
$\Delta m^2_{21}$ [eV$^{-2}$] & $7.50 \times 10^{-5} $ \\ \hline
$\Delta m^2_{31}$ [eV$^{-2}$] & $+2.524 \times 10^{-3}$ \\ \hline
\end{tabular}
\caption{Oscillation parameters used to generate pseudodata for our sterile-neutrino sensitivity analyses for DUNE and Hyper-K. Values taken from Ref.~\cite{Esteban:2016qun}, with the exception of $\delta_{CP}$, which we have taken to be maximally $CP$ violating.}
\label{table:osc_params}
\end{center}
\end{table}

We outline our Monte Carlo simulations of DUNE and Hyper-Kamiokande from Sec.~\ref{sec:experiments}, employing similar procedures to those in Ref.~\cite{Berryman:2015nua} and Ref.~\cite{Kelly:2017kch}, respectively. For DUNE, we assume 3 years of operation each in neutrino and antineutrino modes using the fluxes, efficiencies and detector resolutions reported in Ref.~\cite{Adams:2013qkq}. We assume Hyper-K will run for 2.5 years in neutrino mode and 7.5 years in antineutrino mode, and use the fluxes, efficiencies and detector resolutions reported in Ref.~\cite{Abe:2015zbg}. While Hyper-K is sensitive to atmospheric neutrino oscillations \cite{Kelly:2017kch}, we ignore these in our analysis.

Pseudodata are generated assuming the three-neutrino framework using the experimental specifics mentioned above and the neutrino cross sections in Ref.~\cite{Formaggio:2013kya}. The three-neutrino-oscillations parameters used to generate these pseudodata have been set to their best-fit values in Ref.~\cite{Esteban:2016qun}, shown in Table~\ref{table:osc_params}. The exception is $\delta_{CP}$, which we have taken to be $-\pi/2$; the final sensitivities do not depend strongly on the assumed value. The pseudodata are then analyzed under the four-neutrino hypothesis using the Markov Chain Monte Carlo package \textsc{emcee} \cite{ForemanMackey:2012ig}. We marginalize over the octant of $\theta_{23}$ but assume that the hierarchy will be known by the time DUNE and Hyper-K start collecting data and that the hierarchy is normal; these results do not change significantly for the inverted hierarchy. Gaussian priors are imposed on $\Delta m^2_{21} = (7.50 \pm 0.18) \times 10^{-5}$ eV$^2$ and $|U_{e2}|^2 = 0.299 \pm 0.012$. This analysis differs slightly from those of Refs.~\cite{Berryman:2015nua,Kelly:2017kch} in that $\phi_{24}$ is assumed to be the only nonzero active-sterile mixing angle.

The resulting 95\% C.L. sensitivities in the $\sin^2 2 \theta_{\mu\mu}$ -- $\Delta m^2_{41}$ plane are shown in Figure~\ref{fig:accelerator_data} alongside the constraint from MINOS/MINOS+; all other oscillation parameters have been profiled. The sensitivities of DUNE and Hyper-K for $\phi_{14} = 0$ are not markedly different from those determined in Refs.~\cite{Berryman:2015nua,Kelly:2017kch}. The statistical power of these experiments is dominated by $\nu_\mu$/$\overline{\nu}_\mu$ disappearance, which is sensitive to $\sin^2 2\theta_{\mu\mu}$, whereas the appearance of $\nu_e$/$\overline{\nu}_e$ provides sensitivity to $\sin^2 2\theta_{e\mu}$ (see. Eqs.~\eqref{eq:thaa} and \eqref{eq:thab}). When $\phi_{14}$ vanishes, $\sin^2 2\theta_{e\mu}$ similarly vanishes; there are no sterile-neutrino contributions to $\nu_e$/$\overline{\nu}_e$ appearance. However, this does not impact experimental sensitivity to $\sin^2 2\theta_{\mu\mu}$.

\section{Simulations of CONUS}
\label{app:CONUS}

We provide a brief review of the CE$\nu$NS cross section, as well as some detail on our simulations of CONUS and CONUS100 outlined in Sec.~\ref{sec:experiments}. The cross section for coherent scattering of neutrinos and nuclei is given by \cite{Freedman:1973yd}
\begin{equation}
\label{eq:SMxsec}
\frac{d\sigma}{dT} = \frac{G_F^2 M}{\pi} Q^2_{\rm eff} F^2_{\rm Helm}(q^2) \left(1 - \frac{M T}{2E_\nu^2}\right),
\end{equation}
where $G_F$ is the Fermi constant, $M$ is the mass of the target nucleus, $T$ is the kinetic energy of the recoiling nucleus and $q^2 \approx 2 M T$ is the momentum transferred to the nucleus. In the Standard Model, the effective charge $Q_{\rm eff}$ is given by
\begin{equation}
\label{eq:defineQ}
Q_{\rm eff} = g_V^p Z + g_V^n N,
\end{equation}
where $g_V^n = -\frac{1}{2}$ and $g_V^n = \frac{1}{2} - 2 \sin^2 \theta_W$ are the weak vector charges of the proton and neutron, respectively, with Weinberg angle $\theta_W$. At low energies in the $\overline{\rm MS}$ scheme, $\sin^2 \theta_W \approx 0.23$ \cite{Erler:2004in,Erler:2017knj,Canas:2018rng}, meaning that the weak vector charge of the proton is nearly zero.

The Helm form factor $F_{\rm Helm} (q^2)$ \cite{Helm:1956zz} encodes the distribution of protons and neutrons with the nucleus. This form factor goes to unity when $q^2 \to 0$, and is less than one for finite momentum transfer. For the energies involved in scattering at CONUS, we find that including this form factor gives a $\sim 5\%$ contribution to the cross section. However, searching for sterile neutrinos using CE$\nu$NS requires accounting for these percent-level contributions to the cross section.

We make two simplifying assumptions. The first is that only oscillations related to $\Delta_{41}$ (defined below Eq.~\eqref{eq:osc_probs}) are relevant; given the baseline and neutrino energies available at CONUS, it is reasonable to ignore $\Delta_{31}$ and $\Delta_{21}$. The second assumption is that $U_{e4}$ is nonzero while $U_{\mu 4}$ and $U_{\tau 4}$ vanish. This implies that oscillations depend only $\sin^2 2\theta_{ee}$ (see Eq.~\eqref{eq:thaa}) and that the only nonzero oscillation probabilities are $P_{ee}$ and $P_{es} = 1 - P_{ee}$. Therefore, effects of a sterile neutrino can be included in Eq.~\eqref{eq:SMxsec} by modifying the effective charge $Q_{\rm eff}$ according to
\begin{equation}
\label{eq:steriles}
Q^2_{\rm eff} \to P_{ee} Q^2_{\rm eff};
\end{equation}
only the active component of the neutrino flux at the target will interact via weak neutral currents.

The CONUS experiment is located 17 m from the reactor core at the 3.9 GW$_{\rm th}$ Brokdorf power plant. The detector collects the scintillation light from germanium recoils to observe CE$\nu$NS interactions in the target. The number of events in energy bin $i$ is given by
\begin{widetext}
\begin{equation}
N_i = \Delta t \sum_{f} n_{f} \int_{T_i}^{T_{i}+\Delta T} dT \int_0^{\infty} d E_{\nu} \Phi(E_\nu) \, \frac{d\sigma_f}{dT} \, \Theta(2 E_\nu^2-MT),
\end{equation}
\end{widetext}
where $\Delta t$ is the operating time of the experiment, $f$ represents the five stable\footnote{We note that $^{76}$Ge, while technically unstable, has a long enough half-life to be effectively stable.} isotopes of germanium, $n_f$ is the number of each isotope in the detector, $T_i$ is the lowest energy associated to the bin, $\Delta T$ is the width of the bin, $\Phi(E_\nu)$ is the flux of antineutrinos coming from the reactor, $\frac{d\sigma_f}{dT}$ is the differential CE$\nu$NS cross section involving isotope $f$ and the term $\Theta(2 E_\nu^2-MT)$ enforces the kinematics of the scattering process. 

We use the reactor flux calculation in Ref.~\cite{Kopeikin:2012zz}, normalized to a total antineutrino flux of $2.5 \times 10^{13}$ s$^{-1}$ cm$^{-2}$ \cite{CONUStalk,Farzan:2018gtr}, and include sterile neutrinos following the prescription in Eq.~\eqref{eq:steriles}. The formal upper limit in the integral over $E_\nu$ is infinity, but we cut off the reactor flux above 8 MeV; the flux dies off rapidly above this energy, a feature that has been verified experimentally \cite{NRThesis} and is present in other theoretical calculations of the flux \cite{Mueller:2011nm,Huber:2011wv}. The upper limit in antineutrino energy implies a maximum recoil energy; the masses of the isotopes of germanium imply that this is $\sim$1.75 keV \cite{Farzan:2018gtr}. We take this to be the upper edge of the recoil spectrum for both CONUS and CONUS100.

We form the following $\chi^2$ in order to probe the sensitivity of these experiments to a sterile neutrino, following Ref.~\cite{Farzan:2018gtr}:
\begin{equation}
\label{eq:CONUS}
\chi^2 = \sum_i \frac{\left( N_i^0 - (1 + \alpha) N_i(\sin^2 2\theta_{ee}, \Delta m_{41}^2) \right)^2}{N_i + N_{\rm bkg} + \sigma_f^2 \left( N_i + N_{\rm bkg} \right)^2} + \frac{\alpha^2}{\sigma_\alpha^2},
\end{equation}
where $N^0_i$ is the number of events in bin $i$ with no active-sterile mixing, $N_i(\sin^2 2\theta_{ee}, \Delta m_{41}^2) $ is the same for nontrivial active-sterile mixing, $N_{\rm bkg}$ is the number of background events in each bin, $\alpha$ is a nuisance parameter for the normalization of the flux, $\sigma_\alpha$ is the flux uncertainty and $\sigma_f$ is the uncorrelated shape uncertainty for each bin.

\begin{table}[!t]
\begin{center}
\begin{tabular}{|c||c|c|}\hline
Isotope & CONUS Fraction & CONUS100 Fraction \\ \hline \hline
$^{70}$Ge & 20.5\% & 2.7\% \\ \hline
$^{72}$Ge & 27.4\% & 2.6\% \\ \hline 
$^{73}$Ge & 7.8\% & 1.0\% \\ \hline
$^{74}$Ge & 36.5\% & 4.7\% \\ \hline
$^{76}$Ge & 7.8\% & 88.0\% \\ \hline
\end{tabular}
\caption{The isotopic abundances considered in our simulations of CONUS and CONUS100. Those for CONUS are the natural abundances, while for CONUS100, we assume the target to be 88\% enriched with $^{76}$Ge while the four other isotopes provide the remaining 12\% in proportion to their relative natural abundances.}
\label{tab:germanium}
\end{center}
\end{table}

For our benchmark analysis of CONUS, we take $\Delta t$ = 1 year, and $n_f$ to be consistent with 4.0 kg of natural germanium; the isotopic abundances are shown in Table \ref{tab:germanium}. Because the ionization detection threshold is 0.3 keV and the quenching factor is $\sim 0.25$, the minimum recoil energy is 1.2 keV \cite{Farzan:2018gtr}, and the recoil spectrum is binned in 0.05-keV increments. The normalization and shape uncertainties are taken to be 2\% and 1\%, respectively. 

For CONUS100, we assume a more optimistic experimental configuration. In addition to a five-year run time, the target mass is taken to be 100.0 kg of germanium 88\% enriched with $^{76}$Ge (see Table \ref{tab:germanium}). Furthermore, we assume a recoil threshold of 0.1 keV can be attained and that improvements in reactor antineutrino flux predictions can drive down the normalization and shape uncertainties to 0.5\% and 0.1\%, respectively. For both configurations, we take the background rate to be 1 count/(day$\cdot$keV$\cdot$kg) \cite{Farzan:2018gtr}, even in the low-recoil regime and for the larger target. 

The resulting sensitivity curves are shown in Fig.~\ref{fig:CENNS}. In addition to obvious factors like the larger target mass and improved systematic uncertainties, we highlight two additional sources of improved sensitivity at CONUS100:
\begin{enumerate}
\item The enriched target is primarily $^{76}$Ge. Because the total CE$\nu$NS cross section grows with the (square of the) number of neutrons in the target nucleus (see Eqs.~\eqref{eq:SMxsec} and \eqref{eq:defineQ}), an enriched germanium target yields better statistics relative to a natural germanium target of the same size.
\item The recoil spectrum scales as $\sim\frac{1}{T}$ for low recoils, where $T$ is the kinetic energy of the recoiling nucleus. Lowering the threshold from 1.2 keV to 0.1 keV dramatically increases the number of events at low recoil, where the effects of sterile neutrinos are proportionally more important.
\end{enumerate}

\bibliographystyle{apsrev-title}
\bibliography{cosmo_bib}{}

\end{document}